\documentclass[prb,superscriptaddress,twocolumn,floatfix]{revtex4} 

\usepackage{graphicx}
\usepackage{dcolumn}
\usepackage{bm}
\usepackage[latin1]{inputenc}
\usepackage{amsmath}
\usepackage{amssymb}
\usepackage{graphics}

\usepackage{color}

\begin{document}


\title{Superconducting properties of single-crystalline A$_{x}$Fe$_{2-y}$Se$_{2}$ (A~=~Rb, K) studied using muon spin spectroscopy}

\author{Z.~Shermadini}
\email[Corresponding author:~]{zurab.shermadini@psi.ch}
\affiliation{Laboratory for Muon Spin Spectroscopy, Paul Scherrer
Institute, CH-5232 Villigen PSI, Switzerland}

\author{H.~Luetkens}
\affiliation{Laboratory for Muon Spin Spectroscopy, Paul Scherrer
Institute, CH-5232 Villigen PSI, Switzerland}

\author{R.~Khasanov}
\affiliation{Laboratory for Muon Spin Spectroscopy, Paul Scherrer
Institute, CH-5232 Villigen PSI, Switzerland}

\author{A.~Krzton-Maziopa}
\affiliation{Laboratory for Developments and Methods, Paul Scherrer
Institute, CH-5232 Villigen PSI, Switzerland}

\author{K.~Conder}
\affiliation{Laboratory for Developments and Methods, Paul Scherrer
Institute, CH-5232 Villigen PSI, Switzerland}

\author{E.~Pomjakushina}
\affiliation{Laboratory for Developments and Methods, Paul Scherrer
Institute, CH-5232 Villigen PSI, Switzerland}

\author{H-H.~Klauss}
\affiliation{Institut f\"{u}r Festk\"{o}rperphysik, TU Dresden,
D-01069 Dresden, Germany}

\author{A.~Amato}
\affiliation{Laboratory for Muon Spin Spectroscopy, Paul Scherrer
Institute, CH-5232 Villigen PSI, Switzerland}

\begin{abstract}

We report on the superconducting properties of
A$_{x}$Fe$_{2-y}$Se$_{2}$ (A~=~Rb, K) single crystals studied with
the muon spin relaxation or rotation ($\mu$SR) technique. At low
temperatures, close to 90\% of the sample volumes exhibit
large-moment magnetic order which impedes the investigation of their
superconducting properties by $\mu$SR. On the other hand, about 10\%
of the sample volumes remain paramagnetic and clearly show a
superconducting response. The temperature dependence of the
superconducting carrier density was analyzed within the framework of
a single $s$-wave gap scenario. The zero-temperature values of the
in-plane magnetic penetration depths $\lambda_{ab}(0)$~=~258(2) and
225(2)~nm and the superconducting gaps $\Delta(0)$~=~7.7(2) and
6.3(2) meV have been determined for A~=~Rb and K, respectively. The
microscopic coexistence and/or phase separation of superconductivity
and magnetism is discussed.

\end{abstract}

\maketitle


 The recent discovery of superconductivity in iron selenide
compounds A$_{x}$Fe$_{2-y}$Se$_{2}$ and
(Tl,A)$_{x}$Fe$_{2-y}$Se$_{2}$ (where A = K, Rb, Cs),\cite{Guo,
Maziopa, Wang, Fang} with transition temperatures up to about 32~K,
has led to a renewed interest in iron-based chalcogenide systems.
The average crystal structure of these materials is of the
ThCr$_{2}$Si$_{2}$ type (space group $I4/mmm$).\cite{Rotter} A
remarkable observation is that, beside the superconducting state, a
strong antiferromagnetic state with magnetic moments up to
3.3~$\mu_{\rm B}$ per Fe ion are observed below $T_{\rm N}$ = 478~K,
534~K, and 559 K for A = Cs, Rb, and K, respectively (see
Refs.~\onlinecite{Shermadini, Liu, Bao1, Pomjakushin}). Actually,
the stoichiometry of the parent compound appears to be near
A$_{0.8}$Fe$_{1.6}$Se$_{2}$ (hence the often used denomination
$``245"$). Fe vacancy order has been found to occur below a
structural phase transition $T_{\rm S}$ taking place well above
$T_{\rm N}$.\cite{Bao1, Pomjakushin, Pomjakushin-2011a}

As the interplay with magnetism is thought to play a major role in
understanding the properties of the superconducting state in
iron-based systems, many studies have been devoted to this topic.
The coexistence of magnetism and superconductivity has been reported
in pnictide-``122" systems.\cite{Park, Sanna, Khasanov} A
characteristic of this iron-based family is that the temperature of
the magnetic transition needs to decrease by doping or external
pressure prior to observing a superconducting state at low
temperature. Hence, it appears that in the pnictide-122 systems
static magnetism has to be destroyed by a control parameter such as
doping or pressure before superconductivity can develop its full
strength.\cite{Pratt, Christianson} On the other hand, there are
some indications that the interplay in the chalcogenide iron-based
systems might be rather opposite in nature than the one observed in
the pnictides. Hence, an unusual behavior has been reported in the
FeSe$_{1-x}$ family under pressure, where one observes that both the
magnetic \cite{Bendele} and superconducting
\cite{Bendele,Margadonna} transition temperatures increase with
increasing pressure above 0.8\,GPa.

Hence, as a new iron-based chalcogenide superconductor family, the
A$_{x}$Fe$_{2-y}$Se$_{2}$ systems have attracted many studies
focused on the understanding of the nature of the interplay between
the strong magnetic state occurring at high temperature and the
superconductivity in the same samples. Muon spin relaxation or
rotation $\mu$SR,\cite{Shermadini} transport and
magnetization,\cite{Liu} specific heat, magneto-optical
imaging,\cite{Hu} and  M\"{o}ssbauer \cite{Ryan} spectroscopy
suggest a microscopic coexistence and the bulk character of both the
strong antiferromagnetism and superconductivity. Some studies claim
that superconductivity only occurs in the compositions when Fe
content is compatible with a vacancy order pattern; the ground state
of the material becomes metallic and superconductivity sets
in.\cite{Bao} Alternatively, others suggest that superconductivity
is achieved when the Fe vacancies are disordered and that
superconductivity and magnetism occur in the same samples, but
microscopically separated.\cite{Han} In this Rapid Communication, we
report on $\mu$SR studies specifically devoted to superconducting
properties of the A$_{x}$Fe$_{2-y}$Se$_{2}$ systems, shedding more
light on the question of the bulk character of the superconducting
state at low temperatures.

Single crystals were grown from a melt using the Bridgman
method.\cite{Maziopa} The homogeneity and elemental composition of
cleaved crystals have been studied using x-ray fluorescence
spectroscopy (XRF; Orbis Micro-XRF Analyzer, EDAX), and were
characterized by powder x-ray diffraction using a D8 Advance Bruker
AXS diffractometer with Cu $K_{\alpha}$ radiation. The final
compositions were found to be Rb$_{0.77}$Fe$_{1.61}$Se$_{2}$ and
K$_{0.74}$Fe$_{1.66}$Se$_{2}$. Magnetization and resistivity
measurements have been performed with a physical property
measurement system Quantum Design 9T.

For the $\mu$SR measurements, performed using the transverse-field
(TF) and zero-field (ZF) techniques, the DOLLY instrument located on
the $\pi E1$ beam line of the Swiss Muon Source (Paul Scherrer
Institute, Villigen, Switzerland) was used. Measurements were
performed using a static helium flow cryostat between 2 and 40 K.

\begin{figure}[lt]
\center{\includegraphics[width=1.0\columnwidth,angle=0,clip]{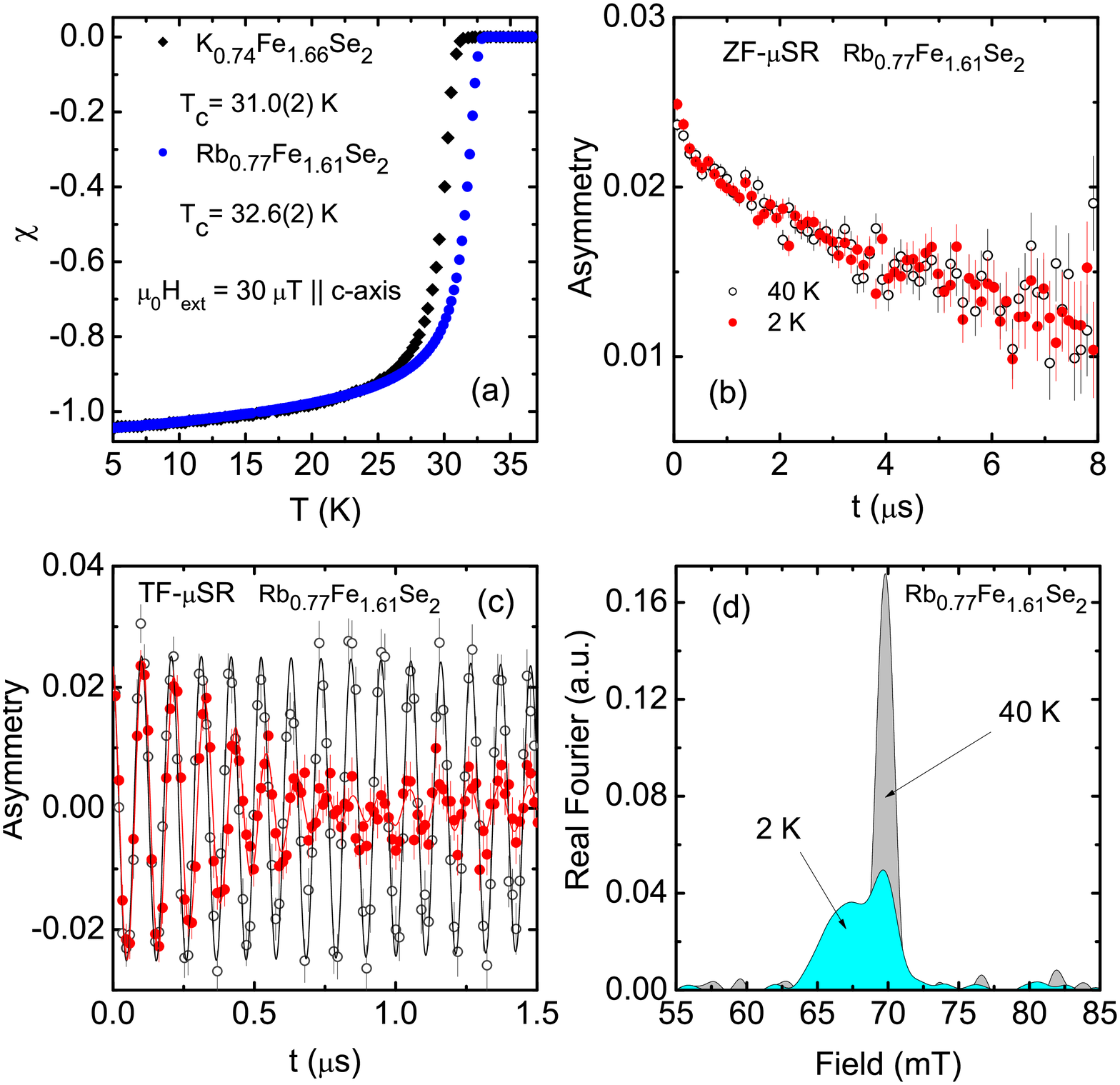}}
  \caption{(a) Temperature dependence of the dc magnetic susceptibility obtained in a zero-field cooling (ZFC) procedure.
  The data were obtained with an external magnetic field of $\mu_{0}$H$_{ext}$ = 30 $\mu $T applied along the $c$-axis.
  (b) Zero field (ZF) and (c) Transverse field (TF) $\mu$SR time spectra recorded above and below $T_{\rm c}$.
  The TF data have been obtained with an external field of 0.07~T and in a field-cooling procedure.
  (d) Fourier transform of the TF $\mu$SR spectra shown on panel (c).}
\label{Asymmetry}
\end{figure}
\begin{figure}[rt]
\center{\includegraphics[width=1.0\columnwidth,angle=0,clip]{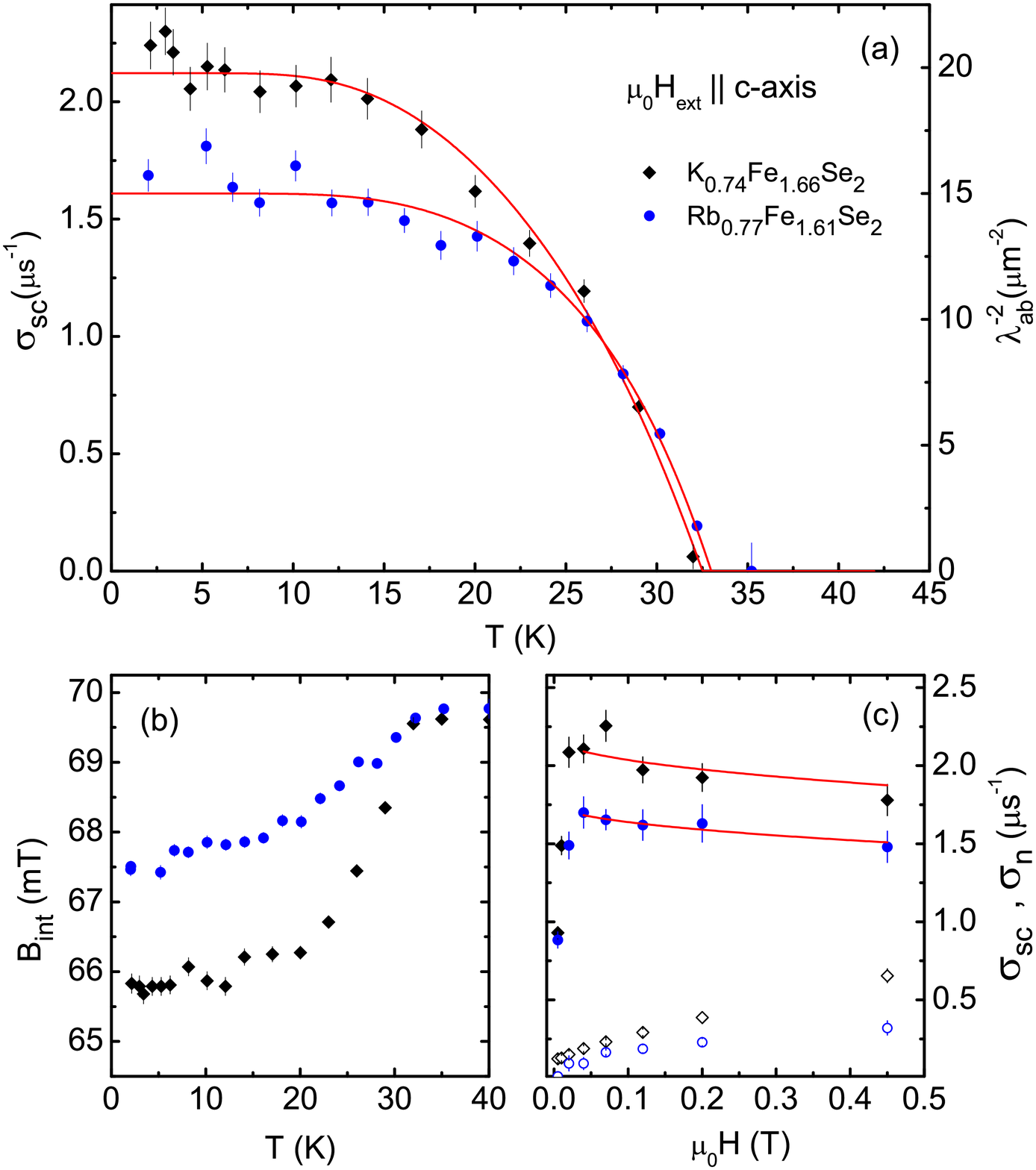}}
  \caption[]{(a) Temperature dependence of $\sigma_{\rm sc} \sim \lambda_{\rm ab}^{-2}$ and therefore of the superfluid density $n_s$ [see
  Eq.(\ref{eq:Lambda}) and text] measured in an applied field of $\mu_{\rm 0}H_{\rm ext}$ = 0.07~T.
   (b) Temperature dependence of the internal field $B_{\rm int}$ sensed by the muons.
   (c) Field dependence of the muon depolarization rate above (40~K; open symbols) and below (2~K; closed symbols) $T_{\rm c}$.
   The external field was applied along the crystallographic $c$ axis.}
   \label{FieldScan}
\end{figure}

A first step of our study has been to elucidate the superconducting
properties of the A$_{x}$Fe$_{2-y}$Se$_{2}$ crystals by performing
in-plane zero-field-cooling (ZFC) magnetization measurements, shown
in Fig.~\ref{Asymmetry}(a). Both samples exhibit sharp
superconducting transitions at $T_c~=~ $31.0(2)~K and 32.6(2)~K for
A = K and Rb, respectively, and a nearly 100\% Meissner screening is
observed. The values of the respective $T_c$ are compatible from the
ones extracted by resistivity (not shown). However, as will be
discussed below, the magnetization study alone is necessary but not
sufficient to claim a 100\% superconducting volume
fraction,\cite{Shen} even though it is very often used that way.

 The first goal of our $\mu$SR study was to check the magnetic properties by the
 ZF and weak TF (wTF) technique. In agreement with our
previous measurements,\cite{Shermadini} we observe that a large
fraction of the $\mu$SR signal is wiped out at very early time
(i.e., $t \ll 0.1~\mu$s), in the wTF as well as in the ZF
measurements, due to a large internal field and/or a broad field
distribution in the antiferromagnetic phase of the sample.
\cite{Shermadini_2} From the wTF measurements it is derived that
this phase represents about 88\% and 89\% of the sample volume for
A~=~Rb and K, respectively. The rest of the signal represents a
fraction of the sample remaining in a paramagnetic state below
$T_{\rm N}$. This sample fraction is characterized by a weak muon
depolarization which is found to be constant between 40~K and 2~K
[see for example the case of A~=~Rb on Fig.~\ref{Asymmetry}(b)].
This temperature independence of the ZF relaxation indicates that
12\% (11\%) of the Rb (K) sample volume is free of a magnetic
transition at least down to 2~K.

This fraction of the sample remaining paramagnetic below $T_{\rm N}$
opens the possibility to study the superconducting state by $\mu$SR,
using the $\mu$SR TF technique.\cite{Sonier} The first step of our
TF $\mu$SR measurements was to determine the optimal external
magnetic field $H_{\rm ext}$ (with $H_{\rm ext} > H_{c1}$) for which
a maximal muon spin depolarization rate ($\sigma_{\rm sc}$, see
below) occurs due to the build-up of a  Flux Line Lattice (FLL) in
the mixed state of the superconductor.\cite{Brandt} The field
dependence of $\sigma_{\rm sc}$ was obtained upon field cooling from
above $T_{\rm c}$ down to 2~K for each data point [see
Fig.~\ref{FieldScan}(c)]. For both Rb and K systems, the optimum
field is above 0.07~T and a complete temperature scan was performed
with this external field applied along the $c$ axis. Typical $\mu$SR
spectra, as well as the corresponding Fourier transforms, are
reported in the panels (c) and (d) of Fig.~\ref{Asymmetry} for
A~=~Rb.

The TF-$\mu$SR time spectra were analyzed using a two-Gaussian
depolarization function:

\begin{eqnarray}
A_{0}P(t)&=& A_{\rm sc}\exp\left(-\frac{(\sigma_{\rm sc}^{2}+\sigma_{\rm n}^{2})t^2}{2}\right)\cos(\gamma_{\mu}B_{\rm int}t+\varphi) \nonumber \\
&&+A_{\rm bg}\exp\left(-\frac{\sigma_{\rm
bg}^{2}t^2}{2}\right)\cos(\gamma_{\mu}B_{\rm bg}t+\varphi)~,
\label{eq:TF}
\end{eqnarray}
where $A_{\rm sc}$ is an initial asymmetry, $B_{\rm int}$ represents
the internal magnetic field at the muon site, and $\sigma_{\rm sc}$
is the Gaussian relaxation rate reflecting the second moment of the
magnetic field distribution due to the FLL in the mixed state.
$\sigma_{\rm n}$, representing the depolarization due to the nuclear
magnetic moments, is taken from the fits above $T_{\rm c}$ and
considered as temperature independent down to 2~K. The second term
of Eq.~(\ref{eq:TF}) represents a background signal (bg)
corresponding to muons stopping in the cryostat walls; $A_{\rm bg}$,
$\sigma_{\rm bg}$ and $B_{\rm bg}$ denote the initial asymmetry
(about 18\% of $A_0$), the relaxation rate and magnetic field (which
has essentially the value of the external field) sensed by muons
stopped in the background.

Due to the very high damping signal occurring in  the
antiferromagnetic phase of the sample \cite{Shermadini} one is
unable to measure any superconducting response for this fraction.
However, as discussed below, this does not exclude that such a phase
presents also a superconducting state.

Figure~\ref{FieldScan}(a) exhibits for both systems the temperature
dependence of the muon depolarization rate $\sigma_{\rm sc}$
reflecting the field distribution created by the FLL. The
temperature dependence of the average value of the internal field
$B_{\rm int}$ sensed by the muon ensemble is reported in
Fig.~\ref{FieldScan}(b). A clear diamagnetic response of the samples
is observed below $T_c$. Considering an extreme-type-II
superconductor, one can evaluate the London magnetic penetration
depth $\lambda$ and superfluid density $n_{s}$ from the second
moment of the magnetic field distribution inside the sample in the
mixed SC state, or alternatively, from the Gaussian muon spin
depolarization rate $\sigma_{\rm sc}$:\cite{Brandt}
\begin{eqnarray}
\frac{\sigma_{\rm sc}^{2}(T)}{\gamma_{\mu}^{2}}&=& 0.00371
\frac{\Phi_{0}^{2}}{\lambda_{\rm ab}^{4}(T)}~,
 \label{eq:Lambda}
 \end{eqnarray}
where $\Phi_{0}$ = 2.068$\times 10^{-15}$~Wb is the magnetic flux
quantum, and $\gamma_{\mu}$/2$\pi$ = 135.5~MHz~T$^{-1}$ is the muon
gyromagnetic ratio (note that as the external field is applied along
the $c$ axis; we are probing the penetration depth $\lambda_{\rm
ab}$ in the basal plane). In turn, from the temperature dependence
of $\lambda_{\rm ab}$, one obtains the temperature evolution of the
superfluid density $n_{s}$ as $n_{s}(T) / n_{s}(0) = \lambda_{\rm
ab}^{-2}(T) / \lambda_{\rm ab}^{-2}(0)$. Here we would like to
mention that the described analysis neglects any additional
contribution to the $\mu$SR relaxation rate due to possible FLL
disorder or induced magnetism.\cite{Sonier2011} Therefore the
extracted value of the penetration depth represents a lower limit.
The temperature dependence of $n_{s}$ was analyzed within the
framework of a BCS single $s$-wave symmetry superconducting gap
$\Delta$.\cite{Tinkham, Wojek} The results of the analysis for
A$_{x}$Fe$_{2-y}$Se$_{2}$ (A~=~K, Rb) are reported in
Fig.~\ref{FieldScan}(a). The solid line represents the fit of a
simple $s$-wave model to the data. Due to the flattening of
$\sigma_{\rm sc}(T)$ below $T_{c}$/2 a clean $d$-wave model is
incompatible with the data. Note that a two gap ($s$+$s$) as well as
an anisotropic $s$-wave scenario provides also a satisfactory
$\chi^{2}$ fitting criteria. The parameters extracted from the
fitting procedure using the simplest $s$-wave model are summarized
in Table~\ref{table1}. The observed values of
$2\Delta(0)/k_{B}T_{c}$ indicate that A$_{x}$Fe$_{2-y}$Se$_{2}$
systems are in the strong-coupling limit.

\begin{table}[tbp]
\caption{List of the parameters obtained from the analysis of the temperature dependence of $n_s$.} 
\centering 
\begin{tabular}{c c c c} 
\hline\hline 
\\
 & Rb$_{0.77}$Fe$_{1.61}$Se$_{2}$ &~K$_{0.74}$Fe$_{1.66}$Se$_{2}$ &~~Unit  \\ [0.7ex]
\hline \\
$T_{\rm c}$ & 32.6(2) & 31.0(2) &~~ K \\
$\lambda_{ab}(0)$ & 258(2) & 225(2) &~~ nm \\ 
$\Delta(0)$  & 7.7(2) & 6.3(2)  &~~~~ meV \\
$2\Delta(0)/k_{B}T_{c}$ & 5.5(2) & 4.7(2)  \\
[2ex] 

\hline 
\end{tabular}
\label{table1} 
\end{table}

The reader should keep in mind that the penetration depth obtained
from the data analysis corresponds to the paramagnetic fraction
representing about 12\% of a total sample volume. We note that NMR
measurements \cite{Torchetti-2011} gave $\lambda=290$~nm, which is
also almost certainly representative for the paramagnetic fraction
only since the NMR signal from the strong antiferromagnetic regions
of the sample is probably wiped out. On the other hand, macroscopic
magnetization\cite{Tsindlekht} and torque\cite{Bosma} measurements
give a considerably longer $\lambda = 580$ and 1800~nm,
respectively, since they probably reflect a kind of average over the
whole sample. Our analysis provides also a slightly lower value of
the superconducting gap than the one measured by the ARPES
technique\cite{Zhang} (isotropic superconducting gap of 10.3~meV).

Figure~\ref{FieldScan}(c) shows the field dependence at 2~K of the
muon depolarization rate obtained upon field cooling from above
$T_{\rm c}$ down to base temperature. Above $\mu_{0}H_{\rm ext}$ =
0.07~T, $\sigma_{\rm sc}$ decreases only very slightly indicating a
high value of the critical field $H_{\rm c2}$. Previous measurements
reported values on the order of $\mu_0H_{\rm c2}^{\rm c}$(0)~=~60~T
for Rb$_{0.88}$Fe$_{1.76}$Se$_{2}$ (Ref.~\onlinecite{Wang}) and for
K$_{0.8}$Fe$_{1.81}$Se$_{2}$ (Ref.~\onlinecite{Mun}). The solid
lines in Fig.~\ref{FieldScan}(c) correspond to a fit based on the
numerical Ginzburg-Landau model (NGL) with the local (London)
approximation ($\lambda \gg \xi$; $\xi$ is the coherence
length)\cite{Brandt} for both systems. This model describes the
magnetic field dependence of the second moment of the field
distribution created by the FLL and therefore the field dependence
of the $\mu$SR depolarization rate. Fixing the value of $\mu_0H_{\rm
c2}(2{\rm K})=55~$T found in the literature\cite{Wang,Mun} and
considering $\lambda_{\rm ab}$ as a free parameter, we get
$\lambda_{\rm ab}$(2K)~=~246(1) and 221(3)~nm for A~=~Rb and K,
respectively. These values are in good agreement with the values
obtained by studying the temperature dependence of the muon
depolarization rate (see Table~\ref{table1}).


 Since the observation of strong magnetism ($m_\mathrm{Fe}
> 2 \mu_\mathrm{B}$ and $T_\mathrm{N}~=~478$~K) \cite{Shermadini} in one of the members of
the newly discovered A$_{x}$Fe$_{2-y}$Se$_{2}$ superconductors, the
most intriguing question to answer by theory as well as by
experimental observation is whether or not superconductivity and
magnetism may coexist microscopically or whether they live apart
together in the same sample but in a phase-separated manner.
Unfortunately experimental techniques that simultaneously can
measure strong magnetism and superconductivity locally are lacking.
Therefore conclusions have to be drawn from a combination of
observations obtained from two or more experimental methods. There
are good arguments for both scenarios. First we will summarize a few
arguments in favor of bulk superconductivity.

In a first step we will discuss techniques that provide macroscopic
information on the superconducting state. In the majority of the
reports on the superconducting properties of the new compounds a
100\% Meissner screening is observed by magnetization measurements,
for a great variety of compounds (see, e.g., Ref.~\onlinecite{Liu}).
Even a decent diamagnetic screening is sometimes observed in
field-cooled magnetization experiments.\cite{Ying-2011} Also a
sizable peak is observed in specific-heat measurements
\cite{Hu,Zeng-2011,Tsurkan} at the superconducting $T_{\rm c}$. A
superconducting volume fraction of 92 - 98\% is estimated from the
specific-heat data by comparing the zero-temperature residual and
the normal-state Sommerfeld coefficients.\cite{Tsurkan} These two
different macroscopic observations in favor of bulk
superconductivity can anyhow be questioned. In samples showing a
100\% Meissner response, anomalies in the magnetic hysteresis loop
were found that can be understood in the picture that
superconductivity in the sample is percolative with weakly coupled
superconducting islands.\cite{Shen} The interpretation of
specific-heat data in view of the superconducting volume fraction is
dangerous since it relies on the determination of the electronic
Sommerfeld coefficient which is assumed to be the same for the whole
sample. This assumption might anyhow not be valid for a potential
phase separation into metallic and insulating volumes. Strong
evidence for bulk superconductivity comes from magneto-optical
imaging\cite{Hu} of a uniform flux distribution after the sample was
cooled in a field which was switched off at low temperatures. This
is consistent with the bulk superconducting nature of the sample and
shows that it is not filamentary or phase separated.\cite{Hu}
Further, different magnetization measurements yield a rather large
$\mu_0H_{c1}=13$~mT and a corresponding magnetic penetration depth
of $\lambda = 580$~nm \cite{Tsindlekht} which is hard to understand
for filamentary superconductivity. On the other hand, in the samples
showing indications of bulk superconductivity, i.e., 100\% Meissner
screening, neutron scattering experiments observe a block spin
antiferromagnetic ordering without traces of a secondary phase,
suggesting a microscopic coexistence of magnetism and
superconductivity.\cite{Bao1,Pomjakushin-2011a, Pomjakushin} Another
argument for a microscopic coexistence comes from two-magnon Raman
scattering.\cite{Zhang-arXiv1106.2706} The intensity of the
two-magnon peak which reflects directly magnetic order undergoes a
clear, steplike reduction on entering the superconducting phase
which suggests a microscopic coexistence of antiferromagnetic order
and superconductivity. Recent inelastic neutron scattering
studies\cite{Park2} observed a magnetic resonant mode below $T_{\rm
c}$ in the Rb$_{2}$Fe$_{4}$Se$_{5}$ system. Such observation also
suggests that bulk SC coexists with $\sqrt{5} \times \sqrt{5}$
magnetic superstructure.

 There are several experiments revealing different kinds of phase separation in different
A$_{x}$Fe$_{2-y}$Se$_{2}$ compounds. Depending on the experimental
technique they are able to directly detect a structural,
non-superconducting/superconducting or a magnetic/nonmagnetic phase
separation. The determination of a magnetic/superconducting phase
separation by these techniques is anyhow only possible on the basis
of plausible arguments. Transmission electron microscopy (TEM)
reveals a rich variety of microstructures related to Fe vacancy
order.\cite{Wang-2011} The superconducting samples clearly appear to
be phase separated suggesting that the superconducting phase could
have a Fe vacancy disordered state. Similarly, scanning nanofocus
single-crystal x-ray diffraction \cite{Ricci} reveals a structural
phase separation in domains with a compressed and an expanded
lattice structure where the latter might be associated with a
magnetic phase adopting a Fe vacancy ordered structure. On the
contrary, scanning tunneling microscopy (STM) able to detect local
structural and electronic properties indicates a microscopic
coexistence of superconductivity and a $\sqrt{2} \times \sqrt{2}$
charge modulation likely caused by block-spin antiferromagnetic
ordering.\cite{Cai} It should be noted anyhow that the STM
measurements did not observe the usually observed Fe vacancy
ordering pattern (which according to neutron measurements exhibits a
block-spin antiferromagnetic state) but rather a vacancy-free FeSe
layer and therefore there might be as well two different magnetic
structures.

Optical spectroscopy observes a Josephson-coupling plasmon of
superconducting condensate.\cite{Yuan} This together with a TEM
analysis suggests a nanoscale stripe-type phase separation between
superconductivity and insulating phases. In addition, optical
conductivity measurements in the THz region observe a very low
charge carrier density in favor of a phase-separated picture with a
minor metallic and a dominant semiconducting phase.\cite{Charnukha}

Local probe techniques such as $\mu$SR (our present and earlier
studies\cite{Shermadini}) and M\"{o}ssbauer\cite{Ksenofontov, Ryan,
Nowik-2011} show a phase separation into a 85--95 \% major magnetic
and a 15--5 \% minor nonmagnetic volume fraction. The paramagnetic
fraction studied by $\mu$SR gives a typical response of
superconducting character. Based on the experimental results one can
suppose that (i) only the antiferromagnetic,  (ii) only the
paramagnetic, or (iii) both regions are superconducting. The
experimental evidence here strongly excludes the first case. On the
other hand, the second scenario is challenged by many experimental
results mentioned above. Unfortunately, the $\mu$SR technique alone
is unable to exclude the second and third scenarios because of a
very high damped muon polarization signal coming from the large
antiferromagnetic fraction. Since both scenarios have their own
experimental support the question remains open and should trigger
further studies of these systems.


In this study we showed that all our A$_{x}$Fe$_{2-y}$Se$_{2}$
samples exhibit a paramagnetic volume fraction of about 12\%. The
$\mu$SR signal of this fraction exhibits a rather weak ZF
depolarization indicating that the paramagnetic islands are rather
large, probably $>$100~nm. The superconducting response obtained by
TF $\mu$SR is typical for a FLL of type-II superconductors again
indicating paramagnetic grains larger than the distance of the flux
lines. The temperature dependence of the superfluid density was
described by a single $s$-wave gap model with zero-temperature
values of the in-plane magnetic penetration depth
$\lambda_{ab}(0)$~=~258(2) and 225(2)~nm and superconducting gaps
$\Delta(0)$~=~7.7(2) and 6.3(2) meV for A~=~Rb and K, respectively.

The $\mu$SR experiments were performed at the Swiss Muon Source,
Paul Scherrer Institut, Villigen, Switzerland. The authors thank the
Sciex-NMS$^{\rm ch}$ (Project Code 10.048) and NCCR MaNEP for
support of this study.

\end{document}